\documentstyle[12pt,fleqn]{article}
\catcode`\@=11
\begin{document} \openup6pt
\title{\Large{Charged Null Fluid Collapse in Anti-de Sitter Spacetimes
and Naked Singularities}}
\author{S.G.~Ghosh\thanks
{email: sgghosh@hotmail.com} \\
\normalsize Department of Mathematics, Science College, Congress Nagar, \\
\normalsize Nagpur-440 012, India}

\vspace{2in}
\date{}
\maketitle

\begin{abstract}
We investigate the occurrence of naked singularities in the
spherically symmetric, plane symmetric and cylindrically symmetric
collapse of charged null fluid in an anti-de Sitter background.
The naked singularities are found to be strong in Tipler's sense
and thus violate the cosmic censorship conjecture, but not hoop
conjecture.
\end{abstract}

\noindent {\bf PACS number(s)}: 04.20.Dw
\\

That the  end state of gravitational collapse of a sufficiently
massive star is a gravitational singularity is a fact established
by the singularity theorems of Hawking and Ellis \cite{he}.
However, these singularities theorems does not guarantee the
existence of an event horizon. The conjecture that such a
singularity from a regular initial surface must always be hidden
behind an event horizon, called cosmic censorship conjecture (CCC)
was proposed by Penrose \cite{rp}. The CCC forbids the existence
of naked singularities.  Despite almost 30 years of effort we are
far from a general proof of CCC (for recent reviews and
references, see \cite{r1}).  But, significant progress has been
made in trying to find counter examples to CCC.  In particular,
the Vaidya \cite{pc} solution that represent an imploding null
fluid with spherical symmetry has been intensively studied for the
formation of naked singularities \cite{ps1}.

Many of studies in the gravitational collapse were motivated by
Thorne's is hoop conjecture \cite{ks} that collapse will yield a
black hole only if a mass $M$ is compressed to a region with
circumference $C \leq 4 \pi M$ in all directions. If hoop
conjecture is true, naked singularities may form if collapse can
yield  $C \geq 4 \pi M$ in some direction.
 Thus, planar or cylindrical matter will not form a black hole (black plane or
black string) \cite{ks}.

However hoop conjecture was given for spacetimes with a zero
cosmological term.  In the presence of negative cosmological term
one can expect the occurrence of major changes.  When a negative
cosmological constant is introduced, the spacetime will become
asymptotically anti-de Sitter spacetime. Indeed, Lemos \cite{jl}
has shown that planar or cylindrical black holes form rather than
naked singularity from gravitational collapse of a planar or
cylindrical matter distribution in an anti-de Sitter spacetime,
violating in this way the hoop conjecture but not CCC.  He also
pointed out that for the spherical case, the collapse proceed to
form naked singularities violating cosmic censorship conjecture,
but not the hoop conjecture.

The purpose of this brief report is to see how the results found
in ref. \cite{jl} get modified for the charged case.
  The usefulness of these models is
that  they do offer opportunity to explore of properties of
singular spacetime and, in the case of curvature singularity to
address issue such as local or global nakedness \cite{gr} and
strength. Such a model may be valuable in attempts to put CCC in
concrete mathematical form.

We find that both spherical and non-spherical collapse of charged
null fluid admit strong curvature naked singularities in
accordance with hoop conjecture and violating CCC.

The charged Vaidya anti-de Sitter metric in
 $(v,r, \theta, \phi)$ coordinates is \cite{cz,ww}
\begin{equation}
ds^2 = - (1 + \alpha^2 r^2 -  \frac{2 m(v)}{r} +
\frac{e^2(v)}{r^2} ) dv^2 + 2 dv dr + r^2 (d \theta^2+ sin^2
\theta d \phi^2), \label{eq:me}
\end{equation}
where $\alpha \equiv \sqrt{- \Lambda/3}$, $\Lambda$ is the
cosmological constant.  $v$ represents advanced Edmonton time, in
which $r$  is decreasing towards the future along a ray $v=const.$
and the two arbitrary functions $m(v)$ and $e(v)$ (which are
restricted only by the energy conditions), represent,
respectively, the mass and electric charge at advanced time $v$.
This metric (\ref{eq:me}) represents a solution to the
Einstein-Maxwell
 equations for a collapsing charged null fluid in the
spherically symmetric anti-de Sitter background.

The model considered here is obtained from energy momentum tensor
of the form
\begin{equation}
T_{ab} = \rho k_{a}k_{b} + T^{(m)}_{ab}, \label{eq:te}
\end{equation}
where $\rho$  in this  case is given by
\begin{equation}
 \rho = \frac{1}{4 \pi r^3} \left[ r \dot{m}(v) - e(v) \dot{e}(v)
\right]  \label{eq:tn}
\end{equation}
with the null vector $k_{a}$ satisfying $k_{a} = - \delta_{a}^{v}
\; and \; k_{a}k^{a} = 0$, $T_{ab}^{(m)}$ is related to the
electro-magnetic tensor $F_{ab}$:
\begin{equation}
T^{(m)}_{ab} = \frac{1}{4 \pi} \left( F_{ac} F_{b}^{c} -
\frac{1}{4} g_{ab} F_{cd} F^{cd} \right) \label{eq:tm}
\end{equation}
which satisfies Maxwell's field equations
\begin{equation}
F_{[a b;c]} = 0 \, and \, F_{a b;c} g^{bc} = 4 \pi J_{a},
\label{eq:mf}
\end{equation}
where $J_{a}$ is the four-current vector.

Clearly, for the weak energy condition to be satisfied we require
the bracketed quantity in Eq. (3) to be non negative.  We note
that the stress tensor in general may not obey the weak energy
condition.  In particular, if $dm/de > 0$ then there always exists
a critical radius $r_{c} = e \dot{e}/ \dot{m}$ such that when $r <
r_{c}$
 the weak energy condition is always violated.  However, in realistic
situations, the particle cannot get into the region $r < r_{c}$
because of the Lorentz force and so the energy condition is still
preserved \cite{ww,oa}.

The Kretschmann scalar ($K = R_{abcd} R^{abcd}$, $R_{abcd}$ is the
Riemann tensor) for the metric (\ref{eq:me}) reduces to
\begin{equation}
K = \frac{48}{r^6} \left[m^2(v) - \frac{2}{r} e^2(v)m(v) +
\frac{7}{6}
 \frac{e^4(v)}{r^2}  \right] + 24 \alpha^4   \label{eq:ks}.
\end{equation}
So the Kretschmann scalar diverges along $r = 0$ for $m$ and $e
\neq 0$, establishing that metric (\ref{eq:me}) is scalar
polynomial singular \cite{he}.

The physical situation is that of a radial influx of charged null
fluid in the region of the anti-de Sitter universe.  The first
shell arrives at $r=0$ at time $v=0$ and the final at $v=T$. A
central singularity of growing mass developed at $r=0$.
  For $ v < 0$ we have $m(v)\;=\;e(v)\;=\;0$, i.e., the anti-de Sitter metric,
 and for $ v > T$,
$\dot{m}(v)\;=\;\dot{e}(v)\;=\;0$, $m(v)\; and \;e^2(v) $ are
positive definite.  The metric for $v=0$ to $v=T$ is  charged
Vaidya-anti-de Sitter, and for $v>T$ we have the Reissner
Nordstr$\ddot{o}$m anti-de Sitter solution.

Radial ($ \theta$ and $ \phi \,=\,const$.) null geodesics of the
metric (1) must satisfy the null condition
\begin{equation}
\frac{dr}{dv} = \frac{1}{2} \left[1 + \alpha^2 r^2 - \frac{2
m(v)}{r} + \frac{e^2(v)}{r^2} \right].         \label{eq:de1}
\end{equation}
Clearly, the above differential equation has a singularity at
$r=0$, $v=0$. The nature (a naked singularity or a black hole) of
the collapsing solutions can be characterized by the existence of
radial null geodesics coming out from the singularity.  The nature
of the singularity can be analyzed by techniques in \cite{r1}. To
proceed further, we choose
\begin{equation}
m(v) = \lambda v \; (\lambda > 0) \;and\; e^2(v) = \mu^2 v^2
(\mu^2 > 0) \label{eq:mv}
\end{equation}
for $0 \leq v \leq T$ \cite{ps1,lz}. Let $y \equiv v/r$ be the
tangent to a possible
 outgoing geodesic from the singularity.
In order to determine the nature of the limiting value of $y$ at
$r=0$, $v=0$ on a singular geodesic, we let
\begin{equation}
y_{0} = \lim_{r \rightarrow 0 \; v\rightarrow 0} y =
\lim_{r\rightarrow 0 \; v\rightarrow 0} \frac{v}{r}.
\label{eq:lm1}
\end{equation}
Using Eqs. (\ref{eq:de1}), (\ref{eq:mv})  and L'H\^{o}pital's rule
we get
\begin{equation}
y_{0} = \lim_{r\rightarrow 0 \; v\rightarrow 0} y =
\lim_{r\rightarrow 0 \; v\rightarrow 0} \frac{v}{r}=
\lim_{r\rightarrow 0 \; v\rightarrow 0} \frac{dv}{dr} = \frac{2}{1
- 2 \lambda y_{0} + \mu^2 y_{0}^2}    \label{eq:lm2}
\end{equation}
which implies,
\begin{equation}
 \mu^2 y_{0}^3 - 2 \lambda y_{0}^2 + y_{0} - 2 = 0.       \label{eq:ae}
\end{equation}
If Eq. (\ref{eq:ae})
 admits one or more positive real roots then the  central shell focusing
singularity is at least locally naked.  Thus the occurrence of
positive roots implies that CCC is violated. In the absence of
positive roots  of (\ref{eq:ae}), the central singularity is not
naked because in that case there are no outgoing future directed
null geodesics from the singularity (for more details, see
\cite{r1}).  Hence when there are no
 positive roots to (\ref{eq:ae}),
the collapse will always lead to a black hole.
 We now examine the condition for the occurrence of a naked singularity.

An interesting point to note from Eq. (\ref{eq:ae}) is that it
admits at least one positive root for $\lambda >0$ and $\mu^2 > 0$
and no negative roots \cite{ac}, e.g.
 Eq. (\ref{eq:ae}) has a positive real $y_0 = 6.26079$ for
 $\lambda = \mu = 1/4$
and $y_0=1.36466$ for $\lambda = 0$, $\mu=1/2$.  It is easy to see
that Eq. (\ref{eq:ae}) admits all three positive roots  if
$\lambda^2 + 18 \lambda \mu^2 \geq 16 \lambda^3 + \mu^2 + 27
\mu^4$. This happens only for $\lambda \leq 0.082$ and $\mu \leq
0.094$.  It is then easy to check that positive roots of  Eq.
(\ref{eq:ae}) $y_0 = 4.53761, 5.74688$ and $9.4686$ corresponds to
$\lambda = 0.08$ and $\mu = 0.09$. Whereas for $\lambda = 0.04, \;
\mu = 0.04$,  the roots are $y_0 = 2.46049, \;16.2218$ and
$31.3177$. For all such values of $y_0$, the singularity will be
naked.
  It follows that the gravitational collapse of a charged null fluid
in an anti-de Sitter background must lead to a naked singularity
regardless of the values of the parameter $(\lambda, \mu)$.

The charged Vaidya metric can be obtained by taking $\alpha = 0$
in Eq. (\ref{eq:me}), however the Eq. (\ref{eq:ae}) remains
unchanged.  Thus the results of collapsing shells
 in anti-de Sitter background are
similar to that of collapsing shells of radiation in Minkowskian
background \cite{lz}, as it should have been expected, since when
$r \rightarrow 0$ the cosmological term $\alpha^2 r^2$ is
negligible.  The global nakedness of singularity can then be seen
by making a junction onto Reissner Nordstr$\ddot{o}$m anti-de
Sitter spacetime.

When $\mu=0$, the metric (\ref{eq:me}) is Vaidya-anti-de Sitter
metric and Eq.
 (\ref{eq:ae}) admit positive roots when $0 < \lambda \leq 1/16$ and hence
singularities  are naked for $0 < \lambda \leq 1/16$ \cite{jl},
which can be shown  gravitationally strong  \cite{gb}.  When
$\mu=0$ and $\alpha = 0$ the singularities are naked again for $0
< \lambda \leq 1/16$ in which case the metric is Vaidya metric
(see \cite{r1}, for a review).

The strength of a singularity is an important issue because there
have been attempts to relate it to stability \cite{djd}. A
singularity is termed gravitationally strong or simply strong, if
it destroys by crushing or stretching any object that falls into
it. Recently, Nolan \cite{bc} gave an alternative approach to
check the nature of singularities without having to integrate the
geodesic equations.  It was shown in \cite{bc} that
 a radial null geodesic which runs into $r=0$ terminates in a  gravitationally
weak singularity if and only if $\dot{r}$ is finite in the limit
as the singularity is approached (this occurs at $k=0$), the
over-dot here indicates differentiation along the geodesics.  So
assuming a weak singularity, we have
\begin{equation}
\dot{r} \sim  d_{0} \hspace{0.2in} r \sim  d_{0} k
\end{equation}
Using the asymptotic relationship above and Eq. (\ref{eq:mv}), the
geodesic equations yield
\begin{equation}
\ddot{v} \sim  - (\lambda y_0 d_0^{-1}k^{-1} - \mu^2 y_{0}^2
d_{0}^{-1} k^{-1} - \frac{\alpha^2}{3}d_{0}k)d_0^2 y_0^2
\label{eq:ddv}
\end{equation}
But this gives
\begin{equation}
\ddot{v} \sim c k^{-1},
\end{equation}
, where $c = (\lambda - \mu^2 y_0)y_0 d_0^{-1}$, which is
inconsistent with
 $\dot{v} \sim  d_{0} y_{0}$, which is finite.  Thus if the coefficient
$c$ of $k^{-1}$ is non-zero, the singularity is gravitationally
strong. This may be false in the case $c=0$, which is equivalent
to $y_0 = \lambda/ \mu^2$.  But inserting this into the root Eq.
(\ref{eq:ae}) gives
\begin{equation}
\mu^2 = \frac{\lambda}{4} \left( 1 \pm (1-8 \lambda)^{1/2}
\right).
\end{equation}
Thus $c=0$ corresponds to a set (in fact a closed curve) of
measure zero in $(\lambda, \mu)$ parameter space and so is not of
physical significance. Therefore, one may say that generically,
the naked singularities is gravitationally strong in the sense of
Tipler \cite{ft}.

Having seen that the naked singularities in our model is a strong
curvature singularity, we check it for scalar polynomial
singularity. The Kretschmann scalar with the help of  Eqs.
(\ref{eq:mv}), takes the form
\begin{equation}
K = \frac{48}{r^4} \left( \lambda^2 y^2 - 2 \lambda \mu^2 y^3 +
\frac{7}{6} \mu^4 y^4 \right)+ 24 \alpha^2 \label{eq:ks1}
\end{equation}
which diverges at the naked singularity and hence the singularity
is a scalar polynomial singularity.

In this section we discuss gravitational collapse of charged null
fluid in plane symmetric and cylindrical symmetric anti-de Sitter
spacetimes.  Let us first consider the case of plane symmetry.
The Einstein-Maxwell equations also have the solution \cite{cz}
\begin{equation}
ds^2 = - (\alpha^2 r^2 -  \frac{2 q m(v)}{r} + \frac{q^2
e^2(v)}{r^2} ) dv^2 + 2 dv dr + \alpha^2 r^2 (dx^2 + dy^2),
\label{eq:pme}
\end{equation}
where $\rho$ in this case is modified as
\begin{equation}
 \rho = \frac{1}{8 \pi r^3} \left[ q r \dot{m}(v) -
\frac{q}{2}e(v) \dot{e}(v)  \right]. \label{eq:ptn}
\end{equation}
Here $- \infty < x, \: y < \infty$ are coordinate which describe
two dimensional zero curvature space which has topology $R\times
R$.  The parameter $q$ has value $2 \pi/ \alpha^2$, is taken from
Arnowitt-Deser-Misner (ADM) mass of corresponding static black
hole found in \cite{cz}. The metric (\ref{eq:pme}) is plane
symmetric Vaidya-like metric representing gravitational collapse
of charge null dust in
 plane symmetric anti-de Sitter spacetime.
Setting $m(v) = const$, $e(v) = 0$ and $\alpha = 0$ one obtains
the Taub metric \cite{at}.  As in the spherically symmetric case,
the physical situation is that of a radial influx of charged null
fluid towards the centre.  The first ray arrives at the centre for
$r=0$, $v=0$, and final shell arrives at $v =T$, say, then can be
matched with exterior static spacetime.  For $v < 0$ we have plane
symmetric anti-de Sitter spacetime.

Since the Kretschmann scalar is given by
\begin{equation}
K = \frac{48 q^2}{r^6} \left[m^2(v) - \frac{2 q}{r} e^2(v)m(v) +
\frac{7 q^2}{6} \frac{e^4(v)}{r^2}  \right] + 24 \alpha^2
 \label{eq:pks}
\end{equation}
, there is scalar polynomial singularity at $r=0$ for $m$ and $e
\neq 0$. As above, further analysis of structure of this
singularity is initiated by study of the radial null geodesics
equation
\begin{equation}
\frac{dr}{dv} = \frac{1}{2} \left[\alpha^2 r^2 - \frac{2 q
m(v)}{r} + \frac{q^2 e^2(v)}{r^2} \right].        \label{eq:pde1}
\end{equation}
Again  Eq. (\ref{eq:pde1}) has a singular point at $r=0$ and $v=0$
and we write as in Eq. (\ref{eq:mv})
\begin{equation}
q m(v) = \lambda v \; (\lambda > 0) \;and\; q^2 e^2(v) = \mu^2 v^2
(\mu^2 > 0) \label{eq:emv}
\end{equation}
and for this case the algebraic equation is
\begin{equation}
 \mu^2 y_{0}^3 - 2 \lambda y_{0}^2 - 2 = 0.       \label{eq:pae}
\end{equation}
 Eq. (\ref{eq:pae}) has atleast one positive roots and no negative
roots (for e.g. a root  $y_0=2.3593$ of Eq. (\ref{eq:pae})
corresponds to $\lambda = \mu= 1$).  For the present case Eq.
(\ref{eq:ddv}) is unaltered and hence singularities are strong
curvature singularities.  Thus referring to above discussion,
collapse lead to a naked singularity irrespective of the values of
the parameter.  The uncharged case can be obtained by taking
$\mu=0$ and Eq. (\ref{eq:pae}) does not admit any positive roots
and hence collapse proceed to form a black hole \cite{jl}.

Finally, we turn our attention to the cylindrical symmetric ant-de
Sitter spacetime.  The metric is
\begin{equation}
ds^2 = - (\alpha^2 r^2 -  \frac{2 q m(v)}{r} + \frac{q^2
e^2(v)}{r^2} ) dv^2 + 2 dv dr + r^2 (d\theta^2 + \alpha^2 dz^2),
\label{eq:sme}
\end{equation}
where $- \infty < v$, $z < \infty $, $0 \leq r < \infty $ and $0
\leq \theta \leq  2\pi$. Here $\rho$ is given by Eq.
(\ref{eq:ptn}), but parameter $q$ has value $2/ \alpha$. The two
dimensional surface has topology of $R \times S^1$.
 Similar to the above discussion is also valid in
this case as well.  Therefore, we can conclude that the
gravitational collapse in the cylindrical symmetric anti-de Sitter
spacetime also forms a strong curvature naked singularities which
are also scalar polynomial.

In this work we have discussed the spherical and the non-spherical
(planar and cylindrical) collapse of charged null fluid in an
anti-de Sitter background.  In the limit $\mu \rightarrow 0$ our
results reduce to those obtained previously \cite{jl}. For the
spherical case, the collapse proceeds in much the same way as in
Minkowskian background \cite{lz} and as in the uncharged case
\cite{jl}, i.e., shell focusing strong curvature naked
singularities do arise violating CCC, but not the hoop conjecture.
Lemos \cite{jl} has shown that non-spherical null fluid
(uncharged) collapse does not yield naked singularities, but
always black holes.  We have shown the inclusion of charge does
alter the result, i.e., non-spherical collapse of charged null
fluid leads to strong curvature singularities.  This shows that in
our non-spherical case also the CCC is violated, but not the hoop
conjecture.

\paragraph{Acknowledgment:} I would like to thank IUCAA,
Pune (India) for kind hospitality while this work was being done.

\noindent


\begin{thebibliography}{99}
\bibitem{ac} First we note that Eq. (\ref{eq:ae}) being a cubic equation,
it must have one real root.  From theory of equation it is easy to
see, for $\lambda > 0$ and $\mu^2 > 0$, any real root of the Eq.
(\ref{eq:ae}) must be positive as negative values of $y_0$ do not
solve this equation.
\end{thebibliography}
\end{document}